\documentclass[aps,pre,onecolumn,11pt,notitlepage]{revtex4-1}
\usepackage{bm,epsfig,graphics,amssymb,amsmath,subeqnarray,setspace,graphicx,amsthm,epstopdf,subfigure,color}
\usepackage{epstopdf, epsf,psfrag, graphicx,color}

\def\b{\mathbf}\def\ep{\varepsilon}

\def\tcb{\textcolor{black}}

\def\b{\boldsymbol}\def\ep{\varepsilon}
\def\a{\bar{a}}
\newcommand{\calM}{{\mathcal M}}

\begin{document}
\title{{Stability and dynamics of magnetocapillary interactions}}
\author{Rujeko Chinomona,\textit{$^{a,b}$} Janelle Lajeunesse,\textit{$^{a}$} William H. Mitchell,\textit{$^{a}$} Yao Yao,\textit{$^{a}$} and Saverio E. Spagnolie\textit{$^{a}$}}
\affiliation{$^{a}$Department of Mathematics, University of Wisconsin-Madison, 480 Lincoln Dr., Madison, WI 53706}
\affiliation{$^{b}$Department of Computational and Applied Mathematics, 6100 Main MS-134, Rice University, Houston, TX 77005}
\date{\today}

\begin{abstract}
Recent experiments have shown that floating ferromagnetic beads, under the influence of an oscillating background magnetic field, can move along a liquid-air interface in a sustained periodic locomotion [Lumay \textit{et al., Soft Matter}, 2013, \textbf{9}, 2420]. Dynamic activity arises from a periodically induced dipole-dipole repulsion between the beads acting in concert with capillary attraction. We investigate analytically and numerically the stability and dynamics of this {\it magnetocapillary swimming}, and explore other related topics including the steady and periodic equilibrium configurations of two and three beads, \tcb{and bead collisions.} The swimming speed and system stability depend on a dimensionless measure of the relative repulsive and attractive forces which we term the magnetocapillary number. An oscillatory magnetic field may stabilize an otherwise unstable collinear configuration, and striking behaviors are observed in fast transitions to and from locomotory states, offering insight into the behavior and self-assembly of interface-bound micro-particles.
\end{abstract}
\maketitle
\section{Introduction}

The last decade has seen a burst of interest in the manipulation of colloidal particles, including applications such as tunable smart materials and micro-scale self-assembly \cite{wg02}, and the behavior of colloidal particles bound to a liquid-air interface and forced by electric and magnetic fields \cite{sakt04,grclw04,asjn08,as08,lvv09,jnfsa09,bgsa10,pssa13,smhg13}. Focuses have included particles of differing types \cite{od08,dsy13,pssa13b}, the dynamics of self-assembled vesicles \cite{vd07}, time-dependent forcing \cite{tfjs08,opdfzb09,dsy13}, Janus particles \cite{ybblg12,kptl13}, self-propelled structures \cite{snezhko11}, optical effects \cite{tdmw11}, the rate of cluster formation \cite{kkp13}, and self-assembly on ultra-soft gels \cite{cc14,cc14b}. The introduction of colloidal building blocks into soft media such as fluid interfaces, nematic liquid crystals, or more complex mesophases creates distortions of the medium which can be used to fabricate more elaborate colloidal objects \cite{wg02,DietrichColloids,SurfactantColloidCrystals,CapillaryInteractionsInclusions,Lapointe2009,ColloidalDispersions}. Much effort has been devoted to achieving complex self-assembly by tailoring the shape of the elementary colloidal building blocks \cite{Lapointe2009,LockKeyColloids}. 

In what is playfully known as the ``Cheerios effect,'' \cite{vm05,DietrichColloids}, identical particles floating at an air-liquid interface experience capillary forces which act to draw them together. The Cheerios effect and similar surface-mediated aggregation have been investigated in the context of vesiculation \cite{MembraneColloidAggregation,MembraneVesiculation}, colloidal flocculation \cite{DietrichColloids,ColloidalDispersions}, millimetric ecology \cite{MeniscusClimb,Voise2011}, and the buckling and folding dynamics of floating filaments \cite{Audoly11,esbl13}. 

Colloidal suspensions, upon the introduction of another physical force such as a magnetic field, can exhibit surprising behaviors and dynamics. Localized magnetic snake and aster shapes can emerge when the colloidal suspension is confined at the interface between two immiscible liquids and is energized by the alternating magnetic field \cite{sa11,pssa13}. Piet et al. provided an experimental and theoretical study of this snake-aster transition, and showed that viscosity can be used to control the outcome of the dynamic self-assembly in magnetic colloidal suspensions \cite{pssa13b}. If the system is well characterized theoretically, the dominant force balance can be tuned to yield desirable shapes. Recent examples include the use of adhesion and delamination \cite{WrinkleFoldTransition,LocalizedBuckling,CapillaryWrinklingMembranes,CompressionIntegrableDaimantWitten,PocivavsekFolding,BrauParametric,EvansLauga2009}, and swelling and capillary interactions \cite{ElastoCapReview,MeniscusLithography,NanopillarAssembly}.
\begin{figure}[htbp]
\begin{center}
\includegraphics[width=.45\textwidth]{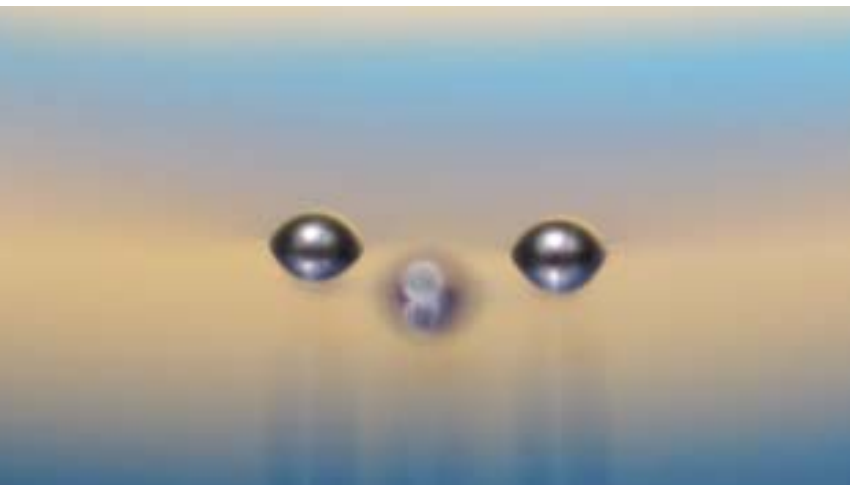}
\caption{Three paramagnetic beads afloat on a meniscus. Reproduced from Lumay et al.~\cite{lowv13}, with permission from The Royal Society of Chemistry.}
\label{Lumay}
\end{center}
\end{figure}
In a recent experiment, Lumay et al.~\cite{lowv13} studied the peculiar dynamics of a small number of identical floating ferromagnetic beads, which in addition to capillary attraction also experience dipole-dipole magnetic repulsion induced by an external magnetic field (see Fig.~\ref{Lumay}). A balance between the attractive capillary forces and the repulsive magnetic forces leads to a self-assembled equilibrium structure. Then, upon the introduction of an oscillating magnetic field in the plane parallel to the fluid surface, the self-assembled structure was found to transport itself through the liquid while undergoing a periodic internal dynamics. Subsequent efforts by the same group showed more intriguing transitions \cite{vctdml12,hgckwov13} and self-assembly of magnetocapillary swimmers \cite{vol13}.

In this paper, we investigate analytically and numerically the equilibrium configurations of two and three beads in a constant and an oscillatory magnetic field, study the stability of these configurations to small perturbations, and explore the {\it magnetocapillary swimming} dynamics described by Lumay et al.~\cite{lowv13}. Swimming speeds and stability properties are determined analytically as a function of a dimensionless measure of the relative repulsive and attractive forces which we term the magnetocapillary number.  The fluid is assumed to be sufficiently viscous, or the beads sufficiently small, so that the Stokes equations of viscous flow apply. For certain physical parameters, an oscillatory magnetic field is shown to stabilize an otherwise unstable collinear configuration, and striking behaviors are observed in fast transitions to and from propulsive states. \tcb{In addition, large oscillatory magnetic fields can induce bead collisions.} The results may provide insight into the behavior and self-assembly of interface-bound micro-particles.

The paper is organized as follows. The equations of motion are described in \S\ref{sec:equations}, along with a description of the dimensionless parameters which characterize the system, and a note on the numerical method used. In \S\ref{sec:equilibrium} we describe equilibrium states of two and more beads in the case of a fixed background magnetic field, and we study analytically the stability of these equilibria with the introduction of an oscillatory component of the magnetic field. In \S\ref{ref:swimming} we choose parameters that lead to a periodic mode of locomotion, and we study the effects of the various parameters on the mean translational velocity. We conclude with a discussion in \S\ref{sec:conclusion}.  

\section{Equations of motion and dimensionless parameters}
\label{sec:equations}

We begin by describing the forces acting on a system of $N$ negatively buoyant colloidal particles confined to an air-liquid interface. In the case of a single floating bead, the equilibrium shape of the interface is determined by a force balance of the effective weight of the bead (the bead weight minus the Archimedean buoyancy force) and surface tension. The length scale over which the surface exhibits significant curvature is the capillary length, $\ell_c=\sqrt{\gamma/\Delta \rho g}$, where $\gamma$ is the interfacial surface tension, $\Delta \rho$ is the density difference between the two fluids, and $g$ is the acceleration due to gravity ($\ell_c \approx 2$mm for an air-water interface) \cite{dbq04}. In the case that two identical beads are floating on the surface, the surface area energy is reduced when they are drawn nearer to each other, giving rise to an attractive force between them (the Cheerios effect). We denote the position of the $i^{th}$ bead center by $\b{x}_i$, the vector from the $i^{th}$ bead to the $j^{th}$ bead by $\b{r}_{ij}=\b{x}_j-\b{x}_i$, the interparticle distance by $r_{ij}=|\b{r}_{ij}|$, and we define $\b{\hat{r}}_{ij}=\b{r}_{ij}/r_{ij}$. The surface deformation due to the presence of the $j^{th}$ bead leads to an attractive force on the $i^{th}$ bead given by
\begin{equation} \label{}
\b{F}^j_c = 2\pi \gamma \a\, \mbox{Bo}^{5/2}\Sigma^2 K_1 (r_{ij}/\ell_c)\b{\hat{r}}_{ij},
\end{equation}
where $\gamma$ is the surface tension, $\a$ is the bead radius, $\mbox{Bo}$ is the Bond number, $\mbox{Bo}=(\a/\ell_c)^2$, and $K_1(\cdot)$ is the first modified Bessel function of the second kind. Finally, $\Sigma = (2\delta - 1)/3 - \cos(\theta_c)/2 +\cos^3(\theta_c)/6$, where $\theta_c$ is the contact angle at the bead-air-fluid interface and $\delta$ is the ratio of the bead and liquid densities \cite{vm05}. 

The introduction of an external magnetic field to the system can lead to induced dipole-dipole repulsion or attraction between floating paramagnetic particles. An isolated paramagnetic bead in a uniform field of strength $\b{H}_0$ induces a magnetic dipole moment $\b{m}$, where
\begin{gather}
\b{m}=\frac{4}{3}\pi \a^3\chi \b{H}_0,
\end{gather} 
and $\chi$ is the effective magnetic susceptibility. The dipole moment associated with each particle is then given by $\b{m}=(4/3)\pi \a^3 \chi \mathbb{B}/\mu_0$, where $\mathbb{B}$ is the local flux density which combines the flux density of the external magnetic field $\mathbb{B}_0$ ($\mathbb{B}_0=\mu_0 \b{H}_0$, where $\mu_0$ is the free-space permeability), and a local dipolar component. The resulting force on the $i^{th}$ bead due to the induced dipole on the $j^{th}$ bead is given by 
\begin{gather} 
\b{F}^j_m = -\frac{3\mu_0}{4\pi} \left(\frac{(\b{m\cdot m})\b{\hat{r}}_{ij} - 5(\b{m} \cdot \b{\hat{r}}_{ij})^2 \b{\hat{r}}_{ij} + 2(\b{m} \cdot \b{\hat{r}}_{ij} ) \b{m}}{r_{ij}^4}\right)
\end{gather}
(see \cite{jackson1962classical}). We will consider external magnetic fields with a constant vertical component (in the $\b{\hat{z}}$ direction, perpendicular to the fluid surface) and an oscillatory horizontal component (in the $\b{\hat{x}}$ direction, parallel to the fluid surface).  This yields $\mathbb{B}=\bar{B}_z\b{\hat{z}}+\bar{B}_x\sin(2\pi \bar{f} t)\b{\hat{x}}$, with $\bar{f}$ the frequency of the oscillating magnetic field, and so the induced magnetic force may be written as
\begin{gather} 
\b{F}^j_m=-\left(\frac{4\pi \a^6\chi^2\bar{B}_z^2}{3\mu_0}\right)\left(\b{F}^j_0+\left(\frac{\bar{B}_x\sin(2\pi \bar{f} t)}{\bar{B}_z}\right)^2\b{F}^j_x \right),
\end{gather}
where
\begin{gather}
\b{F}^j_{0} =-\frac{\b{\hat{r}}_{ij}}{r_{ij}^4},\ \ \ 
\b{F}^j_{x} =\frac{-\b{\hat{r}}_{ij} + 5(\b{\hat{x}} \cdot \b{\hat{r}}_{ij})^2 \b{\hat{r}}_{ij} - 2(\b{\hat{x}} \cdot \b{\hat{r}}_{ij} ) \b{\hat{x}}}{r_{ij}^4}.
\end{gather}

\begin{figure*}
\centering
\includegraphics[width=.95\textwidth]{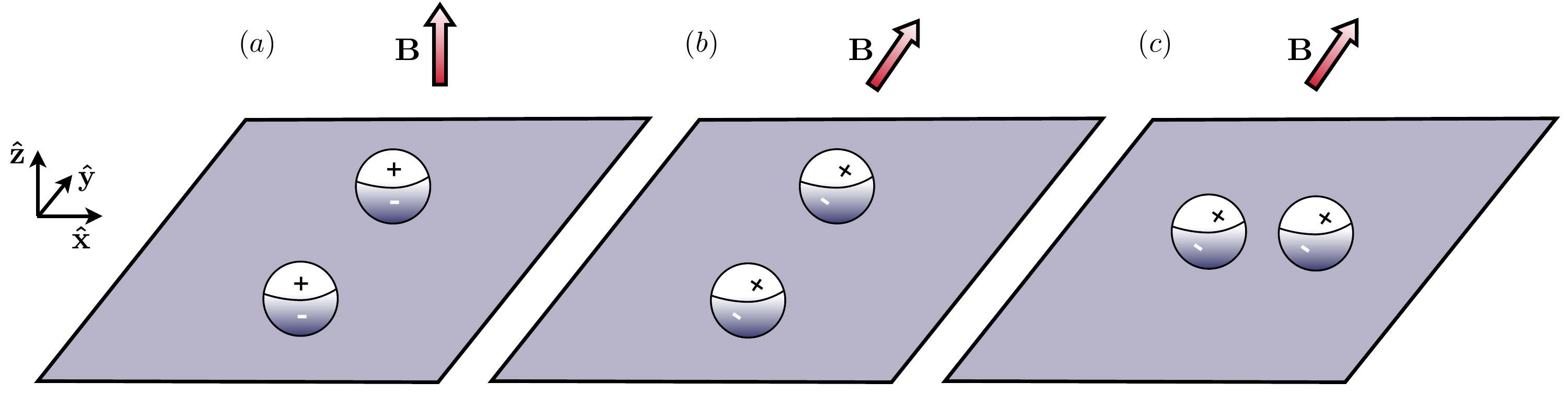}
\caption{Illustration of the anisotropic effect of the horizontal component of the magnetic field on two floating beads. (a) A vertical field leads to an induced dipole-dipole repulsion. (b) A component of the magnetic field in the direction perpendicular to the line of centers between the beads leads to a tilting of the dipole axes, leaving the bead repulsion unchanged for fixed $|\mathbb{B}|$. (c) If the beads are aligned with the horizontal magnetic field component, the tilting of the dipole axes result in a reduced repulsion, or even a magnetic attraction for sufficiently large horizontal field amplitude $B_x$.}
\label{IllustrateB}
\end{figure*}

As illustrated in Fig.~\ref{IllustrateB}, this force can induce further repulsion or attraction, depending on the arrangement of the beads relative to the direction of the magnetic field. If the beads are aligned perpendicular to the horizontal field, the dipole axes will tilt in unison as shown in Fig.~\ref{IllustrateB}b, resulting in a slight extra repulsion between the beads due to an increase in the magnitude of the total magnetic flux density $|\mathbb{B}|$. However, if the beads are aligned parallel to the horizontal field, the magnetic dipole axes will rotate so as to diminish the magnetic attraction between the beads, as shown in Fig.~\ref{IllustrateB}c.

Absent any other forces on the beads it will be shown that locomotion is impossible, so that the viscous hydrodynamic forces must be considered to account for the locomotion seen in the experiments. We will consider the fluid regime where the viscous dissipation overwhelms any inertial effects (Stokes flow) \cite{Batchelor}. The Reynolds number associated with individual bead motion through the fluid is given by $\mbox{Re}=\rho \a d \bar{f}/\mu$, where $\rho$ is the fluid density, $d$ is a characteristic amplitude of bead displacement during oscillation, and $\mu$ is the fluid viscosity. The fluid used in the experiments is a glycerol-water mixture, resulting a characteristic viscosity $\mu=10^{-3} \mbox{Pa}\cdot s$. The beads have radius $\a=250\mu m$, the magnetic field frequency is approximately $\bar{f}=3$Hz, and typical displacement amplitudes are $d=\a/5$, resulting in a small Reynolds number: $\mbox{Re}\approx 0.03$.

The fluid flow $\b{u}$ generated by the motion of the $j^{th}$ bead is modeled through the most slowly decaying fundamental solution of the viscous Stokes equations, the Stokeslet singularity \cite{kk05},
\begin{gather}
\b{u}^j = \frac{1}{8\pi\mu r_{ij}} \left(\mathbb{I}+ \b{\hat{r}}_{ji}\b{\hat{r}}_{ji}^T\right)\cdot \b{F}_h,\label{Stokeslet}
\end{gather}
where $\mathbb{I}$ is the identity operator and $-\b{F}_h$ is the viscous drag on the $j^{th}$ bead. The Stokes drag law in an infinite fluid states that $\b{F}_h=6\pi\mu \a \dot{\b{x}}_j$. This is a rough approximation; the beads are only partially immersed and the flow above is appropriate for a fluid of infinite extent and no boundaries. However, the effect of partial immersion may be understood as a reduction in the effective bead radius $\bar{a}$, and the effect of the shear-free liquid-air interface could be modeled using the method of images \cite{kk05, sl12} which at leading order would introduce a factor of $2$ to the flow above (which may cancel the effect of partial immersion). 

Including all of the forces acting on the beads, and under the assumption of linearity (for instance, assuming that the surface deformation gradient is sufficiently small), the particle motions can be understood as a superposition of pairwise interactions. Momentum balance then gives
\begin{gather}
\dot{\b{x}}_i=\sum_{j\neq i} \left(\b{u}^j+\frac{\b{F}^j}{6\pi\mu \a}\right) ,\label{xdot}
\end{gather}
where $\b{F}^j=\b{F}^j_c+\b{F}^j_m$, the combined capillary attraction and magnetic repulsion. 

\subsection{Nondimensionalization}

The system is made dimensionless by scaling lengths on the capillary length, $\ell_c$, forces on the capillary force $\mathcal{F}$, where $\mathcal{F}=2\pi \gamma \a\, \mbox{Bo}^{5/2} \Sigma^2$, velocities on $\mathcal{F}/(6\pi\mu \a)$, and time on $6\pi\mu \a \ell_c/\mathcal{F}$. The dimensionless velocity of the $i^{th}$ bead (where all variables are now assumed to be dimensionless) then satisfies
\begin{gather}
\dot{\b{x}}_i=\sum_{j\neq i} \left(\b{u}^j+\b{F}^j\right),\label{xdot_nondim}
\end{gather}
where
\begin{gather}
\b{F}^j=F(r_{ij})\b{\hat{r}}_{ij}-\calM_c \left(B_x\sin(f t)\right)^2\left(\frac{\b{\hat{r}}_{ij} - 5(\b{\hat{x}} \cdot \b{\hat{r}}_{ij})^2 \b{\hat{r}}_{ij} + 2(\b{\hat{x}} \cdot \b{\hat{r}}_{ij} ) \b{\hat{x}}}{r_{ij}^4}\right),
\end{gather}
and we have defined the dimensionless horizontal magnetic field amplitude $B_x=\bar{B}_x/\bar{B}_z$ and frequency $f=3\mu \a \ell_c \bar{f}/\mathcal{F}$, and the dimensionless force
\begin{gather}
F(r)=K_1 (r)-\frac{\calM_c}{r^4}.\label{Fr}
\end{gather}
In the above we have introduced a key dimensionless constant $\calM_c$, which we term the magnetocapillary number,
 \begin{gather} 
 \calM_c =\frac{1}{\ell_c^4 \mathcal{F}}\left(\frac{4\pi \a^6 \chi^2B_z^2}{3\mu_0 }\right)=\frac{2 \a^5 \chi^2 \bar{B}_z^2}{3\mu_0 \gamma \mbox{Bo}^{5/2} \Sigma^2\ell_c^4},
 \end{gather}
which compares the relative magnitudes of the repulsive magnetic force and the attractive capillary force. Finally, the dimensionless fluid velocities in Eq.~\eqref{xdot_nondim} are written as
\begin{gather}
\b{u}^j = \frac{3 a}{4 r_{ij}} \left(\dot{\b{x}}_j + (\dot{\b{x}}_j \cdot \b{\hat{r}}_{ji})\b{\hat{r}}_{ji}\right),
\end{gather}
where $a=\a/\ell_c$ is the dimensionless bead radius.

In this paper we will restrict our attention to a regime where the particles and inter-particle distances are smaller than the capillary length, $a \ll 1$, and in the forthcoming analysis we will frequently use an approximation of the first modified Bessel function of the second kind, $K_1(r)\approx 1/r$ for $r \ll 1$ (though in the numerical simulations we need not make such an assumption). In the experiments of Lumay et al.~\cite{lowv13}, the bead radius is approximately $10\%$ of the capillary length ($a\approx 0.1$), though the inter-particle distances are not much smaller than the capillary length. The magnetic field strengths are approximately $\bar{B}_z=2.5\cdot 10^{-3}T$ and $\bar{B}_x=4\cdot 10^{-3}T$, with frequencies on the order of $\bar{f}=1$Hz.

Estimating the magnetocapillary number relevant to the experiments is nontrivial given the large number of physical parameters in its definition. However, the equilibrium configuration gives us a clue. When $B_x=0$, an equilibrium for two beads (and three beads, as we will show in the following section) is achieved when $F(r)=0$, where the capillary attraction exactly balances the magnetic repulsion. With $K_1(r)\approx 1/r$, the equilibrium bead distance is then given by $r=\calM_c^{1/3}$. The equilibrium distances described in the experimental work are on the order of $4a$, resulting in an approximate value of the magnetocapillary number of $\calM_c\approx 0.06$. In addition, this gives a means of estimating $\mathcal{F}$ without measuring $\delta, \theta_c, \Sigma$ and $\gamma$. Namely, using $\bar{a}=2.5\cdot 10^{-4} m$, $\ell_c=2.5\cdot 10^{-3}m$, $\mu_0=4\pi\cdot10^{-7} N/A^2$ (the permeability of the vacuum) and $\chi=3.6$ (the magnßetic susceptibility of chrome steel beads \cite{shah}), we find that $\mathcal{F}\approx 2.8\cdot 10^{-8} N$. Appropriate dimensionless quantities for understanding the experiments are then $a=0.1$, $B_x \in [0,2]$, and $f\approx 2.5$ (using $\bar{f}=1 Hz$). These are the values used in the present work unless otherwise stated.

Finally, the dimensional velocity scale in the experiments for frequency $\bar{f}=3$Hz is estimated to be $\mathcal{U}=6 mm/s$. Average speeds of the center of mass in the experiments were found to be as large as one bead radius per period, corresponding to a dimensionless swimming speed of $U\approx(2.5 \cdot 10^{-3}m)*(3 s^{-1})/\mathcal{U}\approx 1.25$. The capillary wavelength in the experiment is $(2\pi\gamma/\rho \bar{f}^2)^{1/3} \approx 6$cm, i.e. much larger than the particle systems, and we will neglect such effects.

As a way to explore parameter space and to test the validity of the analytical expressions that will be derived in this paper, the system of equations \eqref{xdot_nondim} are integrated numerically. Due to the relative velocity dependence of the Stokeslet flow, the system is implicit in the bead velocities at every moment in time. An explicit specification of the bead velocities is recovered by inverting a $2N\times 2N$ matrix for a given configuration at every moment in time. In some cases the numerical integration is tasked with distinguishing periodic swimming motions from periodic but non-swimming states with high confidence. Time-stepping is achieved using the fourth-order Runge-Kutta method, and any results presented in the remainder of the paper remain robust upon reducing the timestep size. We have verified the accuracy of the method by a convergence study and comparison to exact solutions in the simplest symmetric settings. 

\section{Equilibrium states and stability}
\label{sec:equilibrium}

Consider two floating beads under the influence of a fixed vertical magnetic field in the $\b{\hat{z}}$ direction ($B_x=0$). In equilibrium the beads are motionless, and the inter-particle distances are determined by a balance of the induced dipole-dipole repulsion and the capillary attraction, $F(r)=0$ in Eq.~\eqref{Fr}. Approximating the modified Bessel function for $r \ll 1$ as $K_1(r)\approx 1/r$, this balance is achieved when $1/r={\mathcal M}_c/r^4$, so that the equilibrium distance between the beads is given by $r^*=\mathcal {M}_c^{1/3}$. Naturally, the equilibrium distance increases with the magnetocapillary number (a measure of the relative strength of the magnetic repulsion). Linearizing Eq.~\eqref{xdot_nondim} about this fixed point, the equilibrium configuration is easily found to be asymptotically stable to perturbations.

\begin{figure}[htbp]
\begin{center}
\includegraphics[width=.48\textwidth]{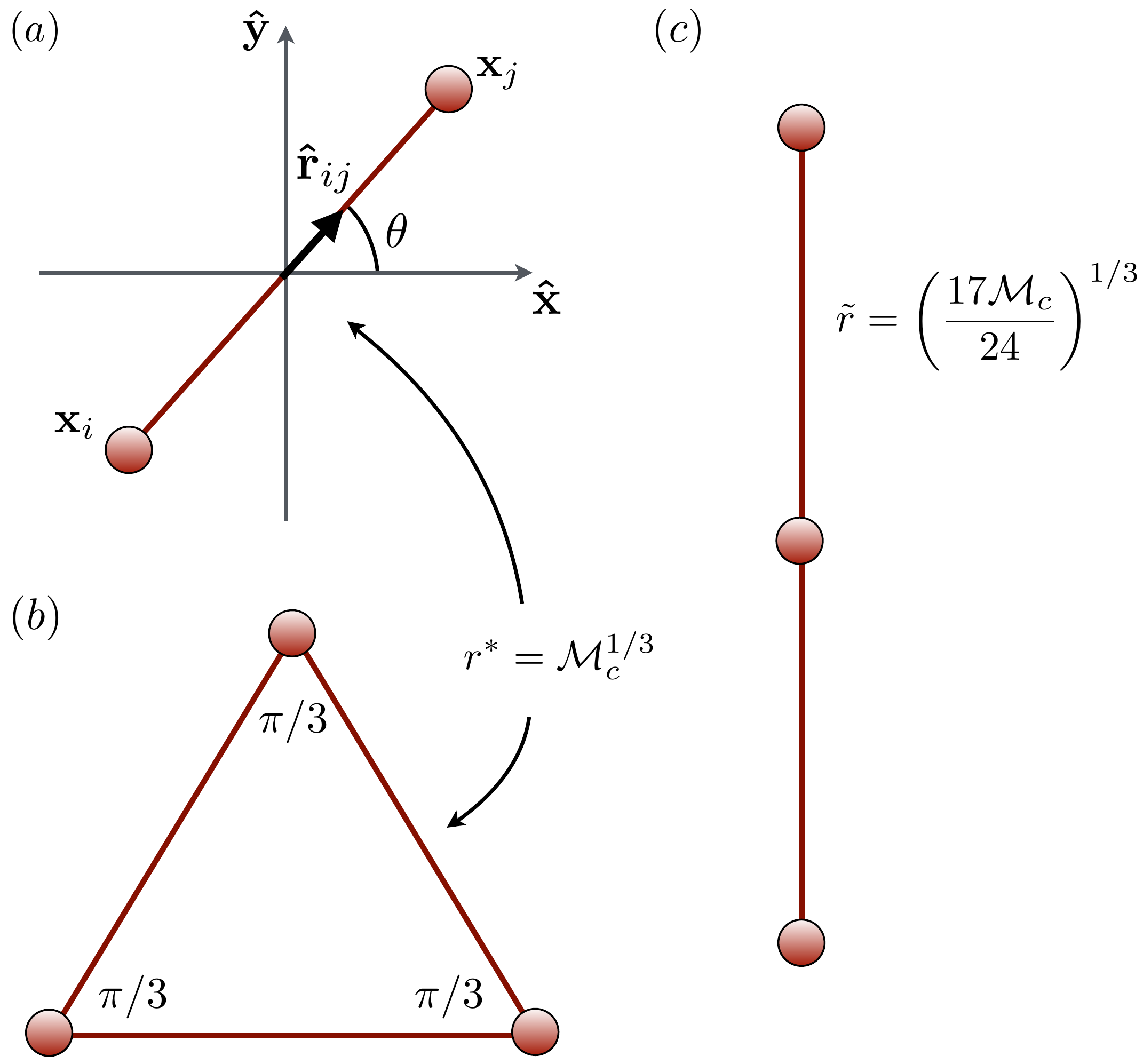}
\caption{Equilibria for (a) two beads and (b,c) three beads for a constant vertical magnetic field. The equilateral triangle is an asymptotically stable configuration, and the inter-particle distances at equilibrium are each identical to the equilibrium distance in the two-bead problem, $r^*=\calM_c^{1/3}$. The equilibrium distance in the collinear case (c) is slightly smaller, $\tilde{r}=(17 \calM_c/24)^{1/3}$.} 
\label{Equilibria}
\end{center}
\end{figure}

Similarly, consider three floating beads located at the vertices of an equilateral triangle as in Fig.~\ref{Equilibria}b. Equilibrium is achieved when each pair $\b{F}^i+\b{F}^j=\b{0}$, for $i,j \in \{1,2,3\}$ and $i\neq j$. Since the beads are not collinear, the three resulting equations require that each pairwise force is balanced, $\b{F}^i=\b{0}$. Hence, the equilibrium distance for three beads in this equilateral placement is identical to that in the two-bead case, $r^*=\mathcal {M}_c^{1/3}$, and just as in the case of two beads this equilateral placement is asymptotically stable. 

Another equilibrium configuration exists when the three beads are collinear, in which case equilibrium is achieved when $F(r)+F(2r)=0$; again using $K_1(r)\approx 1/r$, the beads settle to a slightly smaller distance $\tilde{r}=(17 \calM_c/24)^{1/3}$, as illustrated in Fig.~\ref{Equilibria}c. This collinear conformation is asymptotically stable to perturbations along the axis of the bead placement, but is asymptotically unstable to any perturbations that destroy bead collinearity, with the perturbed configuration rapidly rearranging to form an equilateral triangle. This may also be understood through a simple argument: the inter-particle distance in the three bead case is smaller than in the two bead case, so that neighboring bead pairs are under compression. Given a perturbation that destroys collinearity, the neighboring beads under compression give rise to a force leading to the further degradation of collinearity. The instability of the collinear configuration is reminiscent of the buckling of floating elastic filaments due to capillary self-attraction \cite{esbl13}.

\subsection{Stability of equilibria under an oscillating magnetic field}

The stability of the equilibria illustrated in Fig.~\ref{Equilibria} are straight-forward to analyze, but the introduction of an oscillating component of the magnetic field, $B_x \neq 0$, introduces new complexity and reveals some surprises. In particular, two beads are driven towards alignment with the horizontal field, a sufficiently large oscillation can stabilize the previously unstable collinear configuration, \tcb{and an equilateral configuration is driven towards a perpendicular alignment with the field}. 

\subsubsection{Stability and collision of two oscillating beads } First we consider the simple case of two floating beads. If the beads are aligned with the oscillating magnetic field ($\b{\hat{r}}_{12}\propto\b{\hat{x}}$), the particles will oscillate along the $\b{\hat{x}}$ axis due to both the rotation of the magnetic dipoles and the modulation of the amplitude of the net magnetic field, with the distance between the beads varying in time, and with no movement along $\b{\hat{y}}$ (see Fig.~\ref{IllustrateB}c). Meanwhile, if the beads are aligned perpendicular to the oscillating magnetic field ($\b{\hat{r}}_{12} \propto \b{\hat{y}}$), the particles will move periodically along $\b{\hat{y}}$, now due only to the modulation of the net magnetic field and with no movement along $\b{\hat{x}}$ (see Fig.~\ref{IllustrateB}b). 

More generally, let $\b{\hat{r}}_{12}$ form an angle $\theta$ with $\b{\hat{x}}$ as shown in Fig.~\ref{Equilibria}a, and assume that the beads are placed initially at the equilibrium distance $r^*$ where $F(r^*)=0$. Writing the inter-particle vector as $\b{r}_{12}=\b{x}_2-\b{x}_1=\b{r}=r\left(\cos(\theta)\b{\hat{x}}+\sin(\theta)\b{\hat{y}}\right)$, then using Eq.~\eqref{xdot_nondim} we find
\begin{gather}
\dot{\b{r}}=-\frac{3 a  }{4 r} \left(\dot{\b{r}} + \dot{r}\,\b{\hat{r}}\right)-2F(r)\b{\hat{r}}+\frac{2\calM_c}{r^4} \left(B_x\sin(f t)\right)^2\left((1- 5\cos^2(\theta))\b{\hat{r}} + 2\cos(\theta) \b{\hat{x}}\right).
\end{gather}
Decomposing the system dynamics into its radial and angular components, we have
\begin{gather}
\left(1+\frac{3 a  }{2 r}\right)\dot{r}=-2F(r)-\frac{\calM_c}{r^4} \left(B_x\sin(f t)\right)^2\left(1+3\cos(2\theta)\right),\label{eqr}\\
\left(1+\frac{3 a  }{4 r}\right)\dot{\theta}=-\frac{2\calM_c}{r^5} \left(B_x\sin(f t)\right)^2 \sin(2\theta).\label{eqtheta}
\end{gather}

Assuming a constant separation distance $r$ and $B_x\neq0$, Eq.~\eqref{eqtheta} indicates that the angle $\theta=\pi/2$ (where the beads are set perpendicular to the oscillating field) is unstable, and that the system will be driven to a stable orientation, $\theta=0$ (where the beads are aligned with the oscillating field) for $\theta\in[0,\pi/2)$. In general, however, the distance $r$ also fluctuates in time, but $r$ is uniformly bounded below in the regime of interest (the bodies cannot overlap), which is sufficient to ensure that $\theta=0$ is the only stable equilibrium. Hence, two beads are driven towards alignment with the oscillating horizontal field for any relative amplitude $B_x \neq 0$.

\begin{figure}
\centering
\includegraphics[width=.6\textwidth]{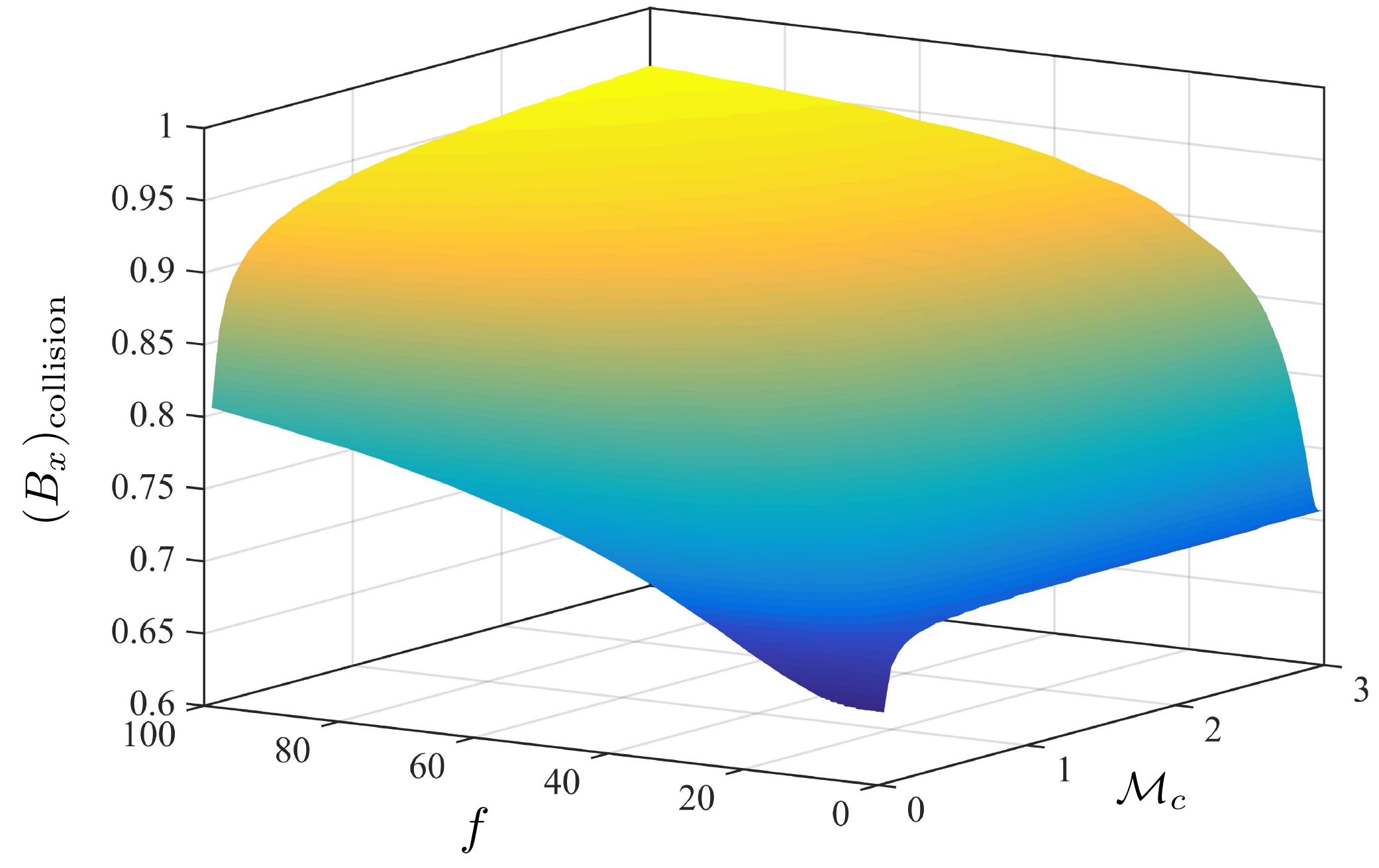}
\caption{The critical value of $B_x$ above which two oscillating beads aligned with the horizontal magnetic field collide.}
\label{Collide}
\end{figure}

\tcb{For a sufficiently large oscillating magnetic field the magnetic dipole moments may undergo a large rotation (see Fig.~\ref{IllustrateB}c) and the two beads may collide. The critical value of $B_x$ above which this occurs, determined by numerical integration of the dynamics, is shown as a function of $\mathcal{M}_c$ and $f$ in Fig.~\ref{Collide}. The value is larger for larger $\mathcal{M}_c$ since the magnetic repulsion is stronger and the beads remain distant, and for larger frequencies when the bead dynamics are driven by the mean horizontal field, $\langle B_x^2\sin^2(ft)\rangle = B_x^2/2$, instead of its full magnitude, $\|B_x^2\sin^2(ft)\|_\infty=B_x^2$.}

\tcb{To estimate this value we set $\theta=0$ in Eq.~\eqref{eqr} and consider two limits. First, in the limit as $f\to 0$, we set $r=2a$ (so that the beads are just in contact) and ask if the interaction is repulsive for all time. With $K_1(r)\approx 1/r$, this results in the critical value $B_x=\sqrt{(1-8a^3\mathcal{M}_c^{-1})/2}$. If $\mathcal{M}_c$ is small, the magnetic repulsion is relatively weak and the beads collide with only a small oscillation of the field (or even with no oscillating field for extremely small $\mathcal{M}_c$), while for any appreciable $\mathcal{M}_c$ the critical value increases towards $B_x=1/\sqrt{2}$, as observed in Fig.~\ref{Collide}. Next, in the limit as $f\to \infty$, ignoring capillary attraction (which can only reduce the critical value of $B_x$ for collision) the beads are certain to collide if the period average of $1-2B_x^2\sin^2(ft)$ is negative, or $B_x=1$. This upper bound is consistent with the numerical results in Fig.~\ref{Collide}.}

\subsubsection{Stability of three oscillating beads } 
\begin{figure*}
\centering
\includegraphics[width=.9\textwidth]{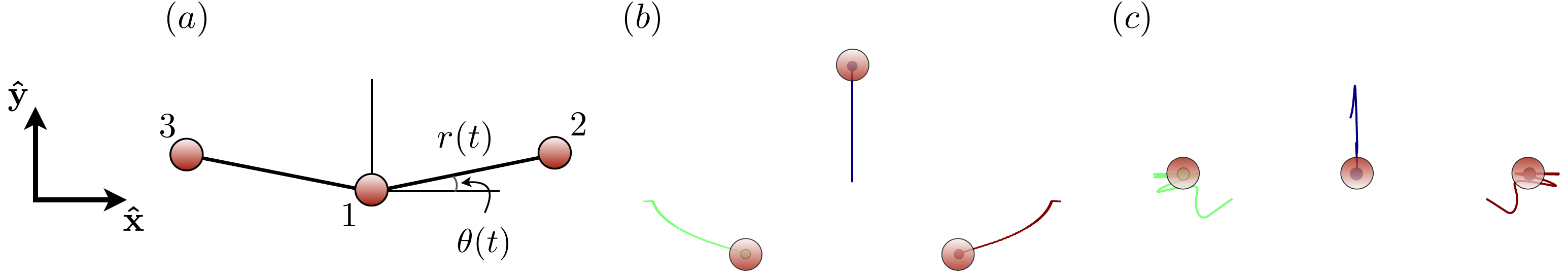}
\caption{(a) A schematic of three beads, symmetrically perturbed. (b) A nearly collinear configuration is unstable for small or zero $B_x$; here with $\calM_c=1.5$ and $B_x=0.25$ the beads are driven towards the nearly equilateral configuration. (c) A large perturbation away from the equilateral configuration with $\calM_c=1.5$ and $B_x=0.7$ is driven to the collinear oscillating state.}
\label{3beads}
\end{figure*}

The stability of three beads due to the oscillating horizontal magnetic field is even more interesting, as there is a surprising stabilization of the collinear state. If the system is initially collinear and placed in alignment with the horizontal magnetic field then the field oscillation acts only to modulate the inter-particle distances in a periodic fashion. However, consider a perturbation to the system that retains up-down symmetry, a translation of the central bead by a relative angle $\theta$ that breaks collinearity as shown in Fig.~\ref{3beads}a. The beads oscillate in time at the frequency of the horizontal magnetic field, but may return towards the oscillating collinear state or may deteriorate towards a different oscillating state. Figure~\ref{3beads}b shows the deterioration of a nearly collinear oscillatory state with $\calM_c=1.5$, $B_x=0.25$, and $\theta(0)=0.1$; the system escapes to a state oscillating about the equilateral triangle configuration. Meanwhile, the bead trajectories using $\calM_c=1.5$ and $B_x=0.7$ are shown in Fig.~\ref{3beads}c, where the system is initialized as a perturbation away from the stable equilateral configuration when $B_x=0$. The system oscillates briefly and then transitions rapidly to the oscillating collinear state. 

\begin{figure}[htbp]
\begin{center}
\includegraphics[width=.6\textwidth]{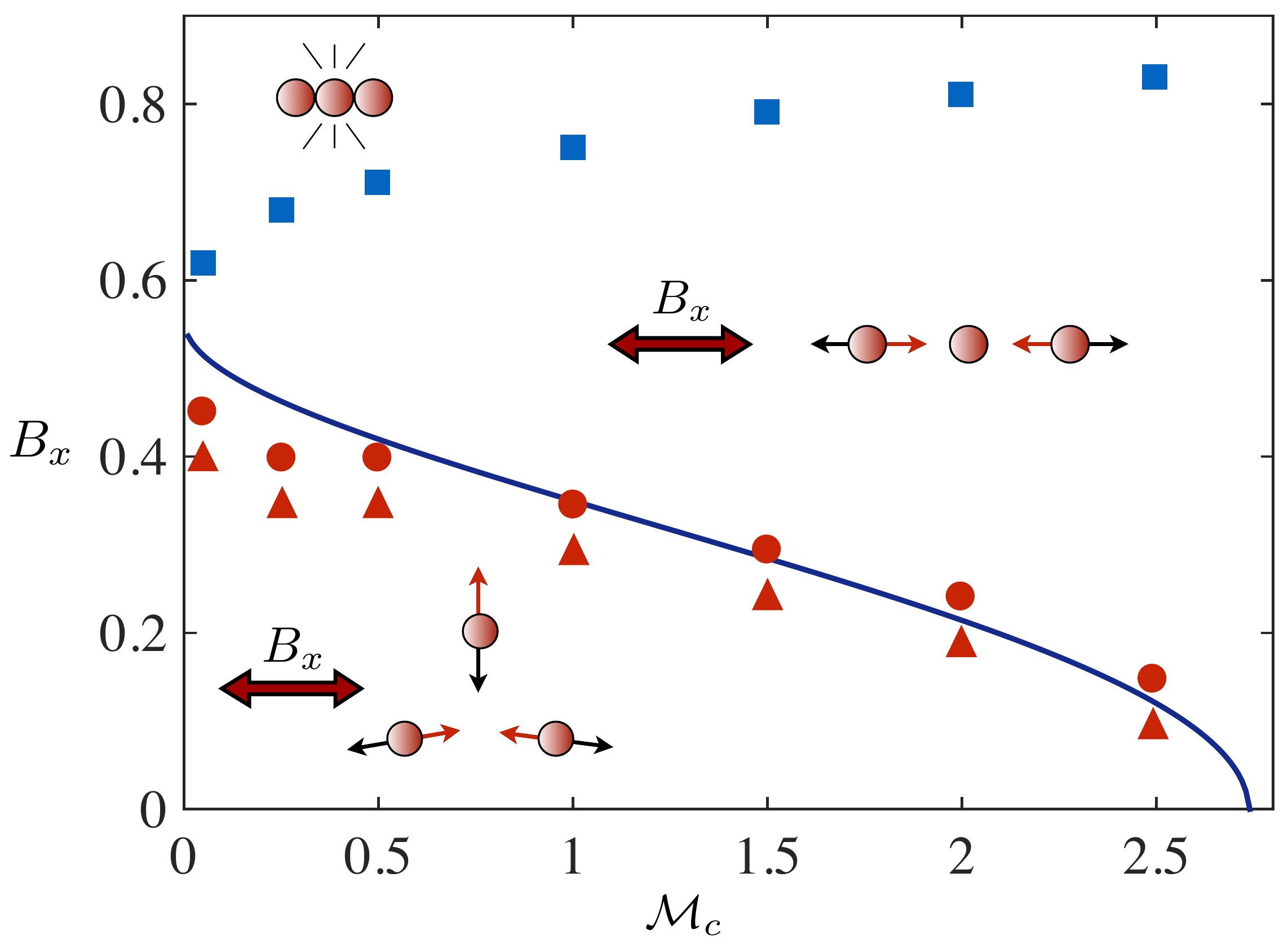}
\caption{The final state of a symmetrically perturbed oscillatory collinear configuration. Symbols show the results of numerical simulations. A symmetrically perturbed collinear system either recovers (circles), or collapses to oscillatory dynamics about the equilateral triangle configuration (triangles). \tcb{Simultaneous motion is indicated by like colored arrows.} The solid line shows the critical value of the horizontal magnetic field strength distinguishing stability/instability as predicted analytically, from Eq.~\eqref{Cx}. The nearly equilateral configuration may still be stable for $B_x>B_x^*$, and in practice we observe hysteresis: the nearly equilateral state is stable to perturbations for $B_x$ much larger than $B_x^*$. \tcb{Square symbols indicate the value of $B_x$ above which the beads collide.}}
\label{Cstar}
\end{center}
\end{figure}

To explore when the system returns to the oscillating collinear state or deteriorates under such a symmetric perturbation we look first to the numerical simulations. For a given value of $\calM_c$ we first determine numerically the (unstable) collinear equilibrium configuration in the case $B_x=0$ (solving for $r^*$), then shift the center bead symmetrically a distance $r^*/10$ to give the initial condition for the oscillatory cases with $B_x> 0$. The system either recovers to an oscillatory collinear state (circles in Fig.~\ref{Cstar}), or collapses to the periodic dynamics about the equilateral triangle configuration (triangles in Fig.~\ref{Cstar}). Beyond a critical value of the horizontal magnetic field strength that depends on the magnetocapillary number, the oscillatory collinear state is found to be stable. The threshold value of the field strength diminishes monotonically to zero with increasing magnetocapillary number, reaching zero at approximately $\calM_c\approx 2.74$. We will return to this special value shortly. \tcb{For yet larger values of $B_x$, the particles again collide. The computed values distinguishing this collapse of the system are shown in Fig.~\ref{Cstar} as squares, and we observe a slow monotonic increase in the critical value with increasing magnetocapillary number as in the two-bead case.}

The critical value of $B_x$ \tcb{which distinguishes the stability of the collinear state to transverse perturbations} may be deduced by studying the equations of motion that describe $r(t)$ and $\theta(t)$ (see Fig.~\ref{3beads}a), and assuming that both $\theta$ and $\calM_c B_x^2$ are small. The hydrodynamic interactions are neglected to simplify the analysis. After dropping all terms of order $O(\theta^2)$ we arrive at the following system, 
\begin{gather}
\dot{r}=-F(r)-F(2r)-\frac{17 C_x}{8 r^4}\sin^2(f  t),\label{eq:drdt}\\
\frac{d}{dt}\ln(\theta)=-\frac{1}{r}\left[2F(r)-F(2 r)+\frac{79 C_x}{8 r^4}\sin^2(f  t)\right],\label{eq:dtheta_dt}
\end{gather}
where we have defined $C_x=\calM_c B_x^2$. To begin our analysis, we denote by $\tilde{r}$ the value of $r$ for which $dr/dt=0$ in \eqref{eq:drdt} when $C_x=0$, or where $F(\tilde{r})+F(2\tilde{r})=0$. Considering first Eq.~\eqref{eq:drdt}, assuming that $C_x \ll 1$ we pursue a regular perturbation expansion that results in the ansatz
\begin{gather}
r(t) = \tilde{r} + C_x\left(A + \frac{g(t)}{f}\right)+O\left(C_x^2\right),
\label{ansatz}
\end{gather}
where $A$ is a constant and $g(t)$ is a mean-zero oscillatory function whose magnitude is of order unity. To find $A$, note that over many periods, the average of the right hand side of Eq.~\eqref{eq:drdt} can be rewritten as $-A\,C_x\left(F'(\tilde{r})+2F'(2\tilde{r})\right)- 17C_x(16 \tilde{r}^4)^{-1} + O(C_x)^2$, and by setting it to 0 we obtain
\begin{equation}
A=-\frac{17 }{16\tilde{r}^4 (F'(\tilde{r})+2F'(2\tilde{r}))}\cdot
\label{eq:a}
\end{equation}

This constant represents a shift in the mean relative position of neighboring beads with the introduction of a nontrivial oscillatory part of the magnetic field. Inserting the ansatz \eqref{ansatz} into Eq.~\eqref{eq:dtheta_dt}, the time-average of the right hand side of Eq.~\eqref{eq:dtheta_dt} over many periods is given by (using $F(\tilde{r})+F(2 \tilde{r})=0$),
\begin{gather}
\frac{1}{\tilde{r}}\Big[-3F(\tilde{r})+C_x \Big(-2F'(\tilde{r}) A + 2F'(2\tilde{r}) A +3A F(\tilde{r})\tilde{r}^{-1}-(79/16)\tilde{r}^{-4}\Big)\Big] + O\left(C_x^2\right).
\end{gather}

The oscillating collinear configuration, $\theta=0$, is stable if and only if this expression is negative, hence by setting it equal to 0 we determine the critical $C_x$, denoted by $C_x^*$, that separates stability from instability. After inserting the value $A$ from above and using $F(\tilde{r})+F(2 \tilde{r})=0$, we find
\begin{gather}\label{Cx}
C_x^*=-\frac{16 \tilde{r}^5 F(\tilde{r}) \left(F'(\tilde{r})+2 F'(2 \tilde{r})\right)}{17F(\tilde{r})+\tilde{r}[15 F'(\tilde{r})+64 F'(2 \tilde{r})]}.
\end{gather}

The critical magnitude of the horizontal magnetic field required for stability, $B_x^*=\sqrt{C_x^*/\calM_c}$ is shown as a continuous function of $\calM_c$ in Fig.~\ref{Cstar}. The approximation, which neglects hydrodynamic interactions and terms of size $O\left(C_x^2\right)$, is reasonable throughout the entire range of $\calM_c$ considered, in part because the resulting values are small and hence are consistent with the omission of the $O\left(C_x^2\right)$ terms. The analysis slightly overestimates the transition value for small $\calM_c$, when the equilibrium distance between beads is small and hydrodynamic interactions become more important. With the magnetocapillary number relevant to the experiments estimated to be approximately $\calM_c=0.06$, the limiting transition value for $B_x^*$ as $\calM_c\rightarrow 0$ of approximately $B_x^*=0.586$ is of note. Based on the nature of its calculation, the approximation Eq.~\eqref{Cx} is likely to be more accurate at higher oscillation frequencies.

The analysis above tells us nothing about the stability of the nearly equilateral configuration, which may still be stable for $B_x>B_x^*$. In fact, in numerical simulations we observe hysteresis: the nearly equilateral state is stable to perturbations for $B_x$ much larger than $B_x^*$. For values $B_x$ just above the critical threshold $B_x^*$, a brief decrease in the magnetic field strength can lead to the destabilization of a nearly collinear state which does not return upon the recovery of the magnetic field.

We pause here to point out some constraints on the physical system. In the calculation above we have used the fact that the beads are repulsive at short distance and attractive at long distance (which is the case for the values $\calM_c$ considered). However, for sufficiently large values of the magnetocapillary number ($\calM_c> 2.74$), $F(r^*)+F(2r^*)=0$ has a solution but the distant beads are no longer attracted to each other; the exponential decay of the Bessel function allows the magnetic repulsion to dominate even in the far-field. The collinear state becomes unstable even to collinear perturbations! Yet another issue arises for $\calM_c>3.34$, as there is no such equilibrium: the beads only interact via dipole-dipole repulsion. In the above and for the remainder of the paper we will assume that $\calM_c<2.74$. 
\begin{figure*}[t]
\centering
\includegraphics[width=.98\textwidth]{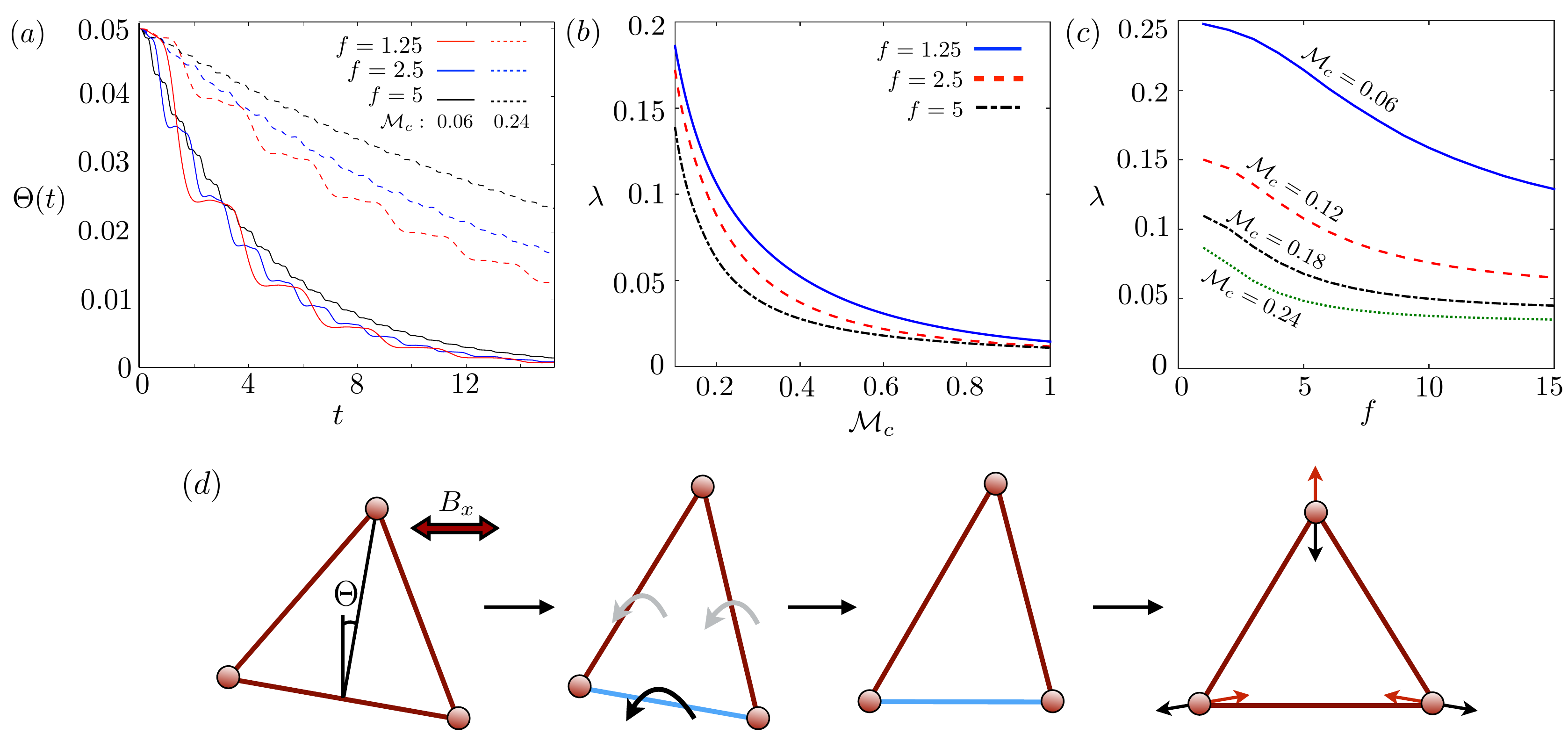}
\caption{\tcb{The oscillating equilateral triangle configuration is stable when oriented perpendicular to the horizontal magnetic field. Here $B_x=0.5$ is fixed. (a) The angle $\Theta(t)$, shown in (d), as a function of time for three frequencies and two magnetocapillary numbers. (b) The decay rate $\lambda$ is shown, where $\Theta(t_n)=\Theta(0)\exp(-\lambda t_n)$ and $t_n=\pi n/f$; $\lambda$ is seen to decay strongly with increasing magnetocapillary number. (c) The decay rate also decreases with increasing frequency of oscillation. (d) The physical mechanism of stability, illustrated. For a small rotation of the stable orientation the horizontal field deforms the configuration into a nearly isosceles triangle; the closest beads rotate strongly towards the stable two-bead alignment with the field, resulting in a return of the system to the unperturbed oscillating configuration. The above may also be seen as a small perturbation away from the symmetrically oriented equilateral case, and hence is also the mechanism by which the symmetric case is unstable to small system rotations.}}   
\label{TriangleStability}
\end{figure*}

\tcb{Returning to the question of collapse, we look to Eq.~\eqref{eq:drdt} and perform the same estimates as for the two-bead case. For $f \to 0$, setting $r=2a$ and asking whether there is a time for which the interaction can be attractive, the resulting estimate is a critical value of the horizontal field, $B_x=\sqrt{(1-192 a^3\mathcal{M}_c^{-1}/17)/2}$, which is very slightly smaller than the estimate in the two-bead case. For large values of $\mathcal{M}_c$ the beads are well separated and a somewhat larger horizontal field amplitude is necessary to induce a collision. Meanwhile, for $f\to \infty$, again ignoring capillary attraction results in an upper bound of $B_x=1$. For finite $f$ and large $\mathcal{M}_c$ the critical value then lies between these two values, as we observe in Fig.~\ref{Cstar}.}

\subsubsection{Stability of the equilateral configuration to small rotations } \tcb{Finally, we observe in simulations that the oscillating equilateral configuration is stable when aligned perpendicular to the magnetic field, (the line of symmetry is in the $\b{\hat{y}}$ direction - the arrangement shown at the bottom left of Fig.~\ref{Cstar}), and is in unstable when aligned symmetrically with the oscillating field. To quantify this stability we define $\Theta$ as the angle between $\b{\hat{y}}$ and $\b{\ell}=\b{x}_1-(\b{x}_2+\b{x}_3)/2$, with $\sin(\Theta)=\b{\hat{x}}\cdot \b{\ell}/|\b{\ell}|$. Fig.~\ref{TriangleStability}a shows $\Theta(t)$ as a function of time for a selection of frequencies and magnetocapillary numbers, with $B_x=0.5$ fixed, and we observe exponential decay with oscillations in every case. The rate of decay, however, depends on the frequency and the magnetocapillary number. We therefore define the exponential decay rate $\lambda$ for the discrete map $\Theta_n=\Theta(t_n)$ with $t_n=\pi n/f$ for integer values of $n$, and $\lambda =-\lim_{n\to \infty} \log(\Theta_n /\Theta_0)/t_n$. Figures~\ref{TriangleStability}b-c show that this decay rate decreases rapidly with increasing magnetocapillary number for $\mathcal{M}_c\in(0,1)$, and also decreases somewhat with increasing frequency.}

\tcb{The physical mechanism underlying this unexpected result is illustrated in Fig.~\ref{TriangleStability}d. For a small rotation of the equilateral configuration the horizontal magnetic field brings the beads on the bottom of Fig.~\ref{TriangleStability}d closer to each other and they rapidly rotate towards alignment with the field (the stable two-bead configuration), resulting after a few cycles to a return of the system to the unperturbed oscillating state. Larger magnetocapillary numbers place the beads further from each other so that the stabilizing two-bead adjustment to the oscillating field is not as strong. Meanwhile, increasing the frequency of oscillation keeps the beads from undergoing large amplitude variations in position, again inhibiting the mechanism, and resulting in smaller decay rates. The mechanism by which the symmetrically oriented equilateral triangle is unstable is identical to the description above - the illustration in Fig.~\ref{TriangleStability}d may be seen instead as a small rotation away from the symmetric state, so that a small rotation drives the system away from the symmetric configuration and towards the stable oscillating state.}


%
%

\section{Fast magnetocapillary swimming requires other physics}
\label{ref:swimming}

In the experiments of Lumay et al.~\cite{lowv13}, it was shown that three or more beads in the presence of an oscillating horizontal magnetic field may swim across the liquid-air interface. To explore the possibility of swimming in the mathematical model studied here, we simulate the dynamics of three beads using the parameters estimated based on the experiments, as discussed in \S\ref{sec:equations}. Namely, we consider a magnetocapillary number $\mathcal M_c = 0.06$, dimensionless frequency $f = 2.5$, and horizontal magnetic field amplitudes $B_x \in [0,1]$. As we have seen in the previous section, for this value of $\calM_c$ and horizontal fields with $B_x \lessapprox 0.5$, the collinear oscillating state is unstable and the system is driven to a nearly equilateral configuration with any perturbation away from collinearity. However, we have also found in numerical simulations that the nearly equilateral configuration is stable for considerably larger values of $B_x$. In particular, simulations suggest that with $\calM_c=0.06$, the nearly equilateral configuration is stable for all values of $B_x$ up to $B_x\approx 0.7$, at which point the magnetic field is large enough to create magnetic attraction and the simulations become nonphysical (the bodies collide).

\begin{figure}
\includegraphics[width=.65\textwidth]{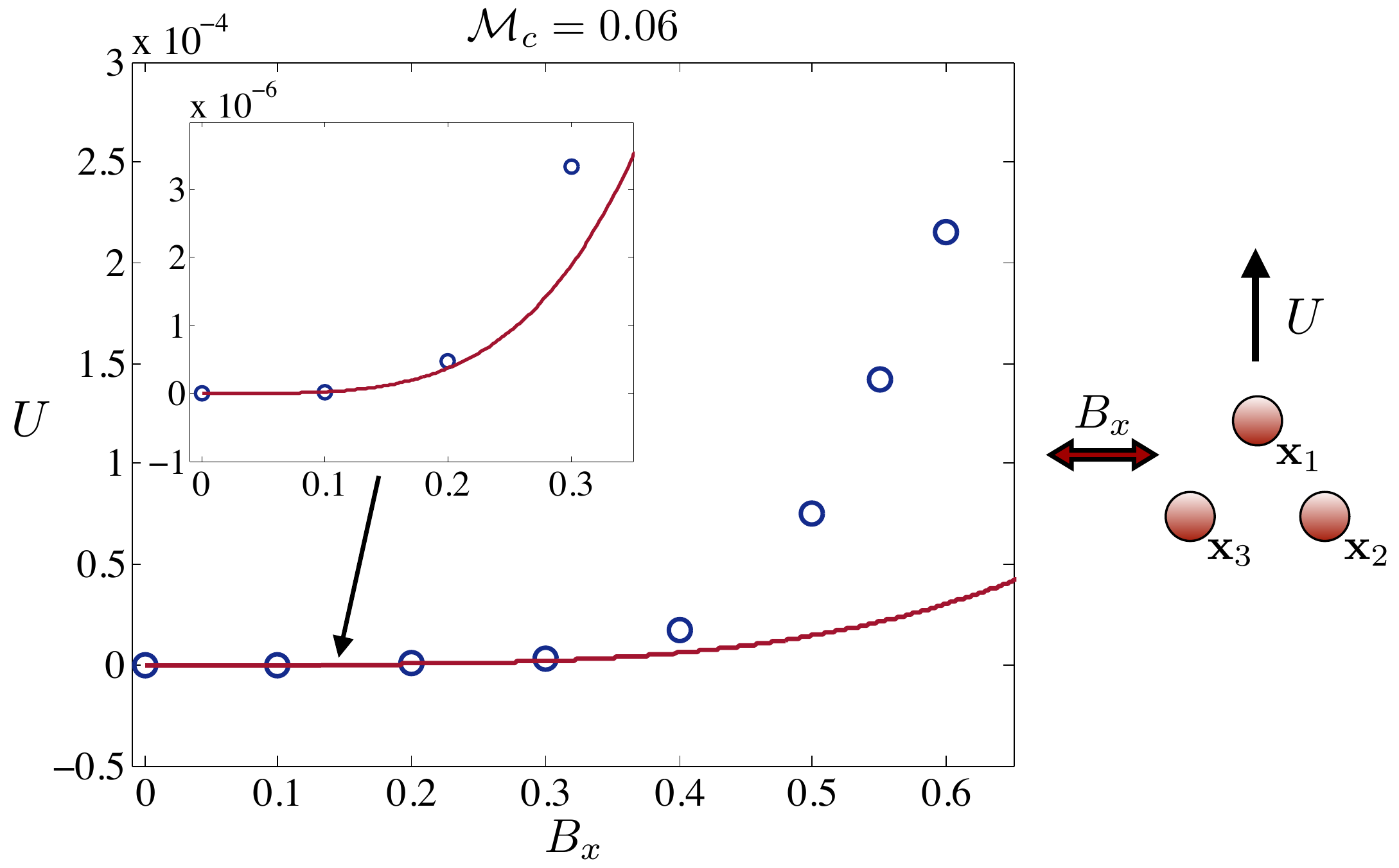}
\caption{The swimming speeds for a range of horizontal magnetic field amplitudes $B_x$ with $\calM_c=0.06$ fixed. The beads swim along the axis of symmetry with monotonically increasing speed with increasing $B_x$ until the beads collide at $B_x\approx 0.7$. Symbols show the result of numerical simulations, while the solid line represents the $O(B_x^4)$ asymptotic approximation from Eq.~\eqref{eq: U2}.}
\label{Unew}
\end{figure}

Setting $B_x=0.5$, and initializing the system in the $B_x=0$ equilateral equilibrium configuration, the three beads oscillate in a periodic and left-right symmetric fashion with a very small net drift along the $\b{\hat{y}}$ direction. Transport in the direction perpendicular to the oscillating part of the magnetic field is consistent with what is observed in the experiments. The system swims along the axis of symmetry with speed $U=1.5\cdot 10^{-4}$, so that each bead translates one bead radius only after many thousands of cycles. Figure~\ref{Unew} shows the computed swimming velocity as symbols for a range of $B_x$, with $\calM_c=0.06$ fixed. The velocity increases monotonically until the bodies collide at $B_x\approx 0.7$. Other simulations not shown here suggest that the swimming mode is stable to small random perturbations of the initial configuration (while large perturbations may deliver the system to the stable oscillatory collinear state for large horizontal magnetic fields $B_x$ as previously described).

We proceed now to investigate the swimming speed by another asymptotic calculation. Once again we consider the case where $|B_x| \ll 1$ and employ a regular asymptotic expansion. The dynamics are constrained to the symmetric configuration illustrated in Fig.~\ref{3beads}b, and the departure of the beads from their relative equilibrium positioning is assumed to be small. Letting $\ep=\calM_c B_x^2 \ll 1$, the bead positions are written as
\begin{gather}
\b{x}_1=\left(\frac{\sqrt{3}}{2}r^*+\ep y^{(1)}_1+... \right)\b{\hat{y}},\label{eq1_0}\\
\b{x}_2=\left(\frac{r^*}{2}+\ep x^{(1)}_2+... \right)\b{\hat{x}}+\left(\ep y^{(1)}_2+... \right)\b{\hat{y}},\\
\b{x}_3=-\left(\frac{r^*}{2}+\ep x^{(1)}_2+... \right)\b{\hat{x}}+\left(\ep y^{(1)}_2+... \right)\b{\hat{y}},\label{eq3_0}
\end{gather}
with $F(r^*)=0$. The expressions above are inserted into the equations of motion, Eq.~\eqref{xdot_nondim}. Taylor expanding about $\ep=0$, and writing $F(r)=F_1(r-r^*)+\dots$, at first order in $\ep$ the resulting system is written compactly as
\begin{gather}
\dot{\b{x}}=  F_1  \mathbb{A} \b{x} + c_B \mathbb{B}  \dot{\b{x}} + \left(1-\cos(2f t)\right)\b{c},
\label{eq:ode}
\end{gather}
where $\b{x} = (y_1^{(1)}, x_2^{(1)}, y_2^{(1)})^T$, $F_1>0$ and $c_B= 3a/(4r^*)$ with 
\begin{gather}
\mathbb{A}=\begin{pmatrix}
  -3/2 & -\sqrt{3}/2 & 3/2 \\
  - \sqrt{3}/4 & -9/4 & \sqrt{3}/4  \\
   3/4  &  \sqrt{3}/4  & -3/4
 \end{pmatrix},\\
 \mathbb{B} = \begin{pmatrix}
 0& -\sqrt{3}/2 &7/2\\
 -\sqrt{3}/4 & -2 & 0 \\
 7/4  & 0 & 1
 \end{pmatrix},
 \label{def:AB}
 \end{gather}
and $\b{c}= [16 (r^*)^4]^{-1}(-2\sqrt{3}, -9,\sqrt{3})^T$. Although the expressions above do not rely on this approximation, should we set $K_1(r)\approx 1/r$ then $r^*=\calM_c^{1/3}$ and $F_1=3/\calM_c^{2/3}$. Inverting to isolate the velocity,
\begin{equation}
\dot{\b{x}} = F_1 (\mathbb{I}-c_B \mathbb{B})^{-1} \mathbb{A} \b{x} +  (1-\cos(2ft)) (\mathbb{I}- c_B \mathbb{B})^{-1}  \b{c}.
\label{eq:ode_2}
\end{equation}
Since $\mathbb{A}$ has one zero eigenvalue and two negative eigenvalues, the same is true of $(\mathbb{I} - c_B \mathbb{B})^{-1} \mathbb{A}$ as long as $c_B$ is sufficiently small. The system is diagonalized by writing $(\mathbb{I} - c_B \mathbb{B})^{-1} \mathbb{A}=\mathbb{T} \mathbb{J} \mathbb{T}^{-1},$ where $\mathbb{J}$ is a diagonal matrix with diagonal terms $0, \lambda_2, \lambda_3$, with $\lambda_2, \lambda_3 < 0$. Note that the first column of $\mathbb{T}$ is $(1,0,1)^{T}$, which is the eigenvector of $\mathbb{A}$ corresponding to the eigenvalue $0$. Performing a change of variables, $\b{y} = \mathbb{T}^{-1} \b{x}$, we have that 
\begin{equation}
\dot{\b{y}} = F_1 \mathbb{J} \b{y} + (1-\cos(2ft)) \b{d},
\label{eq:ode_y}
\end{equation}
with $\b{d}=  \mathbb{T}^{-1}(\mathbb{I}- c_B \mathbb{B})^{-1} \b{c}$, and it is possible to show that the solution must be of the form $\b{y}=\b{e}_0+\b{e}_1\cos(2ft)+\b{e}_2\sin(2ft)$ with constant unknown vectors $\b{e}_0$, $\b{e}_1$, and $\b{e}_2$. In particular, due to the structure of the vector $\b{c}$ there is no linearly growing part of $\b{y}$ in time (see Appendix A). Since we must have $\mathbb{J}\b{e}_0+\b{d}=\b{0}$, the first component of the vector $\b{e}_0$ is left unspecified, so that $\b{y}$ is only determined up to a constant multiple of $(1,0,1)^{T}$, which simply indicates the invariance of the dynamics under translations in the $\b{\hat{y}}$ direction. The remaining vectors $\b{e}_1$ and $\b{e}_2$ are solved by inverting a $6\times 6$ system of equations. The symbolic computation software {\it Mathematica} was used to solve for the somewhat long expressions which we do not reproduce here. The main result of the calculation above, however, is that there is no swimming of order $\varepsilon$ (i.e.~of order $\calM_c B_x^2$). 

To find a nontrivial swimming speed we must proceed with the asymptotic expansion to terms of size $O(\varepsilon^2)=O(\calM_c^2 B_x^4)$. The linear system at the next order is identical to the one at first order but with a more involved forcing term that depends on the $O(\varepsilon)$ dynamics. The full solution at second order is found in a similar calculation as for the solution at first order, but the dynamics are found to involve a term that grows linearly in time. The resulting expression for the swimming speed, writing $U=\varepsilon^2 U_2+O(\varepsilon^3)=U_2\calM_c^2B_x^4$, is found to be
\begin{gather}\label{eq: U2}
U_2=864 \sqrt{3} a f^2 F_1 \left(9 a^2+15 a r^*-8 (r^*)^2\right)\Phi^{-1},
\end{gather}
\begin{multline}
\Phi=(r^*)^5 (9 a-4 r^*) \left(f^2 (15 a+8 r^*)^2+144 F_1^2 (r^*)^2\right)\times\\
\Big(f^2 \left(171 a^2+72 a r^*-64(r^*)^2\right)^2+144 F_1^2 (r^*)^2 (9 a-4 r^*)^2\Big).
\end{multline}
The analytical estimate of the swimming speed is shown in Fig.~\ref{Unew} as a solid line. The approximation begins to deteriorate near $B_x\approx 0.4$. Investigating Eq.~\eqref{eq: U2} allows for the optimization of parameters for maximizing the swimming speed in some contexts. For instance, setting $\calM_c=0.06$ and $a=0.1$, the swimming speed is maximized by selecting the frequency $f\approx 15.2$, which increases $U_2$ to $0.32$ (from $U_2=0.065$ in the case $f=2.5$). Nevertheless, the swimming speed in that case is still exceptionally small. Note that larger oscillation frequencies can increase the available relative horizontal field strengths $B_x$, since the dipole moments are redirected before particles can collide. At least up until the regime of bead collision, with increased $B_x$ comes larger particle excursion distances, larger hydrodynamic interactions, and greater swimming speeds. 

In zero Reynolds number locomotion, the Scallop theorem states that no propulsion is possible if the kinematics are time-reversible (so-named for the single degree of freedom available to a simple scallop) \cite{Purcell77,lp09,Lauga11}. In the present setting, the dynamics of the three beads are very nearly but not quite time-reversible so locomotion is possible. However, the beads nearly move back and forth along the same curve throughout each cycle, so that the kinematics are not sufficiently well removed from reversible dynamics to result in a significant swimming speed. Other simple swimming bodies in viscous fluids are designed specifically to avoid such reversible kinematics, such as the three-bead model swimmer of Najafi and Golestanian \cite{sg04}.

\section{Discussion}
\label{sec:conclusion}

We have investigated the stability and dynamics of two and three floating paramagnetic beads under the influence of capillary attraction and magnetic repulsion. The introduction of an oscillating horizontal magnetic field was found to influence the stability properties in surprising fashion: two beads are driven towards alignment with the oscillatory part of the field, while three beads are driven towards a nearly equilateral arrangement unless the horizontal field is of large amplitude, in which case the horizontal collinear state is asymptotically stable. We proceeded to study the locomotion of the nearly equilateral configuration and compared the results of analytical and numerical calculations, which matched closely for a wide range of $B_x$.

One of the main findings of this research is that the Stokesian hydrodynamic interactions are not sufficient to describe the observed dynamics in the experiments, and to our knowledge the question of precisely how the swimming speeds observed in the experiments are achieved remains open. Although not presented here we have also considered the effect of particle and fluid inertia by solving the Basset-Boussinesq-Oseen equation of unsteady flow \cite{pozrikidis2011introduction}. However, since the Reynolds numbers relevant to the experiments are very small, $\text{Re}\approx0.03$, we have found that inertia has only a very small effect on the computed swimming speed. 

\tcb{We have also investigated the role of Brownian fluctuations, but the dimensionless diffusion constant relevant to the experiments (from the fluctuation-dissipation theorem) leads to forces six orders of magnitude smaller than the capillary and magnetic forces on the beads. Moreover, even for artificially inflated values of the diffusion constant the dynamics are still driven on average by a deterministic swimming motion at the same mean swimming speed. Also recall that the oscillating magnetic field renders the swimming configuration stable to rotational perturbations. Hence, thermal effects which act to reorient the body are over-damped and the swimming system tends to remains on course as it moves through the fluid until the diffusion constant is extremely large. This said, the physical forces that lead to swimming speeds in the work of Lumay et al.~\cite{lowv13} that are orders of magnitude larger than those derived in this paper are almost certainly immune to the effects of thermal fluctuations.}

Other physics appears to be necessary in order to describe the experiments. For instance, we have neglected bead rotations, and their subsequent effects on magnetic dipole moments, inter-particle forces, surface effects, and associated fluid structures and viscous stresses. We have also neglected the partial immersion of the beads, though while viscous drag might seem to reduce the swimming speed, in fact the beads cannot translate at all on average without the hydrodynamic interactions. Finally, the model of capillary attraction may be too simplistic: a modulation of the surface shape and associated forces with bead translation and rotation may be important. These considerations may be examined in a future work.

We are grateful to Nicolas Vandewalle for helpful comments. This work was a product of an NSF-funded summer Research Experiences for Undergraduates (REU) at the University of Wisconsin-Madison, grant number DMS-1056327, with thanks to Andrej Zlatos.

\appendix
 
\section{No magnetocapillary swimming at $O\left(B_x^2\right).$}

Here we show that the three bead system described in \S\ref{ref:swimming} does not swim at first order in $\calM_c B_x^2$, which is assumed to be small. Specifically, we will show that with $\mathbb{A}$ given in \eqref{def:AB}, that the center of mass grows linearly in time as $t\to\infty$ if and only if $c_1+2c_3 \neq 0$, and since this is not the case, there is no swimming at first order. Writing $\b{x}=(y_1^{(1)},x_2^{(1)},y_2^{(1)})^T$, and recalling the change of variables $\b{y}=\mathbb{T}^{-1}\b{x}$, the center of mass at first order is given by 
\begin{gather}
\begin{split}
&\bar{y}(t)=\frac{1}{3}y_1(t) + \frac{2}{3}y_2(t) =  \left(1/3,2/3,0\right) \mathbb{T} \b{y}(t)\\
&= \left(1/3,2/3,0\right)\begin{pmatrix}1 & * &*\\ 0 & * & *\\ 1 & * & * \end{pmatrix} \b{y}(t) = y_1(t) + C_2 y_2(t) +  C_3 y_3(t)
\end{split}\label{swimming_eq}
\end{gather}
for some constants $C_2$ and $C_3$. Now recall that $\b{y}$ satisfies 
\begin{gather}
\dot{\b{y}} = F_1 \mathbb{J} \b{y} + (1-\cos(2ft)) \b{d},
\label{eq:ode_yapp}
\end{gather}
where $\b{d}=  \mathbb{T}^{-1}(\mathbb{I}- c_B \mathbb{B})^{-1} \b{c}$, and $\b{c}= [16 (r^*)^4]^{-1}(-2\sqrt{3}, -9,\sqrt{3})^T$. Component-wise, we write $\dot{y}_i= F_1 \lambda_i y_i +(1-g(t)) d_i$, with $g(t)=\cos(2ft)$. Since $\lambda_2, \lambda_3 <0$, $y_2(t)$ and $y_3(t)$ both remain uniformly bounded in time. As for $y_1(t)$, since the first diagonal term of $\mathbb{J}$ is 0, $y_1(t)$ satisfies $\dot{y}_1= (1-g(t))d_1,$ hence $y_1(t)-d_1 t \left(= d_1 \int_0^{t} g(s) ds + y_1(0)\right) $ is uniformly bounded in time. Using Eq.~\eqref{swimming_eq}, we have that $|\bar{y}(t) -d_1 t|$ is uniformly bounded in time; in other words, the swimming speed is given by $d_1$.

However, $d_1 \neq 0$ if and only if $c_1+2c_3 \neq 0$, which we will now show. Let us denote the columns of $\mathbb{A}$ by $\{\b{a}^1, \b{a}^2, \b{a}^3\}$. By inspection we have that $\b{c} \in \text{span}\{\b{a}^1,  \b{a}^2, \b{a}^3\}$ if and only if $c_1+2c_3=0$. Recall from the diagonalization process that $\mathbb{T}^{-1}(\mathbb{I} -c_B \mathbb{B})^{-1} \mathbb{A} = \mathbb{J} \mathbb{T}^{-1}$, whose first row is a zero vector. In other words, if $c_1+2c_3=0$, then the first component of $\mathbb{T}^{-1}(\mathbb{I}-c_B  \mathbb{B})^{-1} \b{c}$ is $0$ (since $\b{c} \in  \text{span}\{\b{a}^1,  \b{a}^2, \b{a}^3\}$), which gives that $d_1=0$. On the other hand, if $\b{c} \not \in   \text{span}\{\b{a}^1,  \b{a}^2, \b{a}^3\}$, then $d_1$ cannot be 0, since the set $\b{c}$ such that $d_1=0$ has dimension 2 and $\text{span}\{\b{a}^1,  \b{a}^2, \b{a}^3\}$ is already of dimension 2. Inspecting $\b{c}$ we note that indeed $c_1+2c_3=0$, and hence there is no swimming of order $\varepsilon$ (i.e.~of order $\calM_c B_x^2$). 

\bibliographystyle{unsrt}
\bibliography{bibmcs}
\end{document}